\RequirePackage{fix-cm}
\documentclass{svjour3}                     
\smartqed  
\usepackage{graphicx}
\usepackage{amsmath}
\usepackage{amssymb}

\journalname{Ricerche di Matematica}
\begin{document}

 \title{Symmetric form for the hyperbolic-parabolic system \\ of fourth-gradient fluid model
}

\titlerunning{Symmetric form for the system of fourth-gradient
    fluid model}

\author{Henri Gouin and Tommaso Ruggeri} 

\institute{H. Gouin \at Aix-Marseille Univ, CNRS, Centrale Marseille,
    M2P2 UMR 7340,
    13451 Marseille, France \\
    \email{henri.gouin@univ-amu.fr; henri.gouin@yahoo.fr}
    \and
    T. Ruggeri \at
    Department of Mathematics and  Alma Mater Research Center on Applied Mathematics -- AM$^{2}$, University of Bologna, Via Saragozza 8, 40123 Bologna, Italy  \\
    \email{tommaso.ruggeri@unibo.it}  }         %

\date{Received: 7 November 2016 / Accepted: January 2017}

\maketitle

\begin{abstract}
 The fourth-gradient model for fluids - associated with an
  extended molecular mean-field theory of capillarity - is considered.
  By producing fluctuations of density near the critical point like in computational molecular dynamics, the model
  is    more realistic and richer than van der Waals' one and other models associated with a second order expansion.\newline
The aim of the paper is to prove - with a fourth-gradient internal energy already obtained by the mean field theory -  that the quasi-linear system of conservation laws  can be written in an Hermitian symmetric form  implying the stability  of constant solutions. The result extends the  symmetric  hyperbolicity property of governing-equations' systems when an equation of energy associated with  high order deformation of a continuum medium is taken into account.
\end{abstract}\\



\section{Introduction}

Many physical models are represented by quasi-linear first order systems of $%
N$ balance laws (in particular conservation laws),
\begin{equation}
\frac{\partial\mathbf{F}^0(\mathbf{u})}{\partial t}+\frac{\partial%
\mathbf{F} ^{j}(\mathbf{u})}{\partial x^{j}}=\mathbf{f(u)},
\label{sh}
\end{equation}
with an additional scalar balance equation (typically the energy equation in
pure mechanical case or the entropy equation in thermodynamics):
\begin{equation*}
\frac{\partial{h}^0(\mathbf{u})}{\partial t}+\frac{\partial h ^{j}(%
\mathbf{u})}{\partial x^{j}}=  {\it{\Sigma}} (\mathbf{u}),  \label{shsu}
\end{equation*}
where $\mathbf{F}^0,\mathbf{F}^j \left( j=1,2,\dots,n \right)$, $%
\mathbf{f}, \mathbf{u}$ are column vectors of $R^N$ and  ${h}^0$, $h ^{j}$,  $\left( j=1,2,\dots,n \right)$, ${\it\Sigma}$ are scalar functions; $t$, $%
\mathbf{x}\equiv$ $(x^{1},\cdots ,x^{n})$ are the time and space
coordinates, respectively; we adopt sum convection on the repeated indices.%
\newline
Function $h^0$ is assumed convex with respect to field $%
 \mathbf{F}^0 (\mathbf{u})\equiv \mathbf{u}$, \cite%
{Godunov,Friedrichs,Boillat,Ruggeri-1981,Ruggeri2}. Boillat \cite{Boillat} introduces
dual-vector field $\mathbf{u}^\prime$, associated with Legendre
transform $h^{\prime 0}$ and potentials $h^{\prime j}$, such that
\begin{equation}  \label{mainfield}
\mathbf{u}^{\prime}= \left(\frac{\partial h^0}{\partial \mathbf{u}}\right)^\star, \qquad
h^{\prime 0} = \mathbf{u}^{\prime\star} \, \mathbf{u}- h^0, \qquad
h^{\prime j} = \mathbf{u}^{\prime\star}\, \mathbf{F}^j(\mathbf{u})- h^j,
\end{equation}
 where superscript $^{"\star"}$ denotes the
transposition. Therefore by convexity argument, it is possible to take $\mathbf{u}^{\prime}$ as field and we obtain from (\ref{mainfield}):
\begin{equation}  \label{change}
\mathbf{u} =\left( \frac{\partial h^{\prime 0}}{\partial \mathbf{u}^\prime}\right)^\star
, \qquad \mathbf{F}^j(\mathbf{u})= \left(\frac{\partial h^{\prime j}}{%
\partial \mathbf{u}^\prime}\right)^\star.
\end{equation}
Inserting (\ref{change}) into  (\ref{sh}), system (\ref{sh}) becomes
symmetric :
\begin{equation}  \label{symform}
\frac{\partial}{\partial t}\left(\frac{\partial h^{\prime 0}}{\partial
\mathbf{u}^\prime}\right) + \frac{\partial}{\partial x^j}\left(\frac{%
\partial h^{\prime j}}{\partial \mathbf{u}^\prime}\right) = \mathbf{f%
}(\mathbf{u}^{\prime}),
\end{equation}
which is equivalent to
\begin{equation}
\mathbf{A}^0\,\frac{\partial \mathbf{u}^\prime}{\partial t}+%
\mathbf{A}^{j}\frac{ \partial \mathbf{u}^\prime}{\partial x^{j}}=%
\mathbf{f}(\mathbf{u}^\prime),  \label{symm}
\end{equation}%
where matrix $\mathbf{A}^0\equiv \left(\mathbf{A}^0\right)^\star  $ is symmetric
positive  definite and matrices $\mathbf{A}^{j}=\left(\mathbf{A}%
^{j}\right)^{\star }$ are symmetric  :
\begin{equation}  \label{matrici}
\mathbf{A}^{0}\equiv \left(\mathbf{A}^0\right)^\star= \frac{\partial^2
h^{\prime 0}}{\partial \mathbf{u}^{\prime}\partial \mathbf{u}%
^{\prime}}, \qquad \mathbf{A}^j\equiv \left(\mathbf{A}%
^{j}\right)^{\star }= \frac{%
\partial^2 h^{\prime j}}{\partial \mathbf{u}^{\prime}\partial
\mathbf{u}^{\prime}}, \quad (j=1,2,\dots,n).
\end{equation}
The symmetric form of governing equations implies hyperbolicity. For
conservation laws with vanishing productions, the hyperbolicity is
equivalent to the stability of constant solutions with respect to
perturbations in form $\ e^{i(\mathbf{k}^{\star }\mathbf{x}-\omega
t)} $,\ where $i^{2}=-1,\ \mathbf{k}^{\star }=[k_{1},\cdots ,k_{n}] \in (R^n)^\star $ and
$\omega$ is a real scalar. Indeed, the symmetric form of governing equations
for an unknown vector $\mathbf{u}, \ (\mathbf{u}^\star = [u_1,\cdots
,u_{n}]$ implies the \emph{dispersion relation :}
\begin{equation*}
{\rm{det}}\,(\mathbf{A}_{(k)}-\omega \mathbf{A}^0)=0 \quad \mathrm{%
with} \quad \mathbf{A}_{(k)}=\mathbf{A}^{j}k_{j}\,,  \label{eigenval}
\end{equation*}%
which determines real values of $\omega $ for any \emph{real wave vector\,}$%
\mathbf{k}$. In this case, phase velocities are real and coincide with
the characteristic velocities of hyperbolic system \cite{RMSeccia,BLR}.
Moreover right eigenvectors of $\mathbf{A}_{(k)}$ with respect to $%
\mathbf{A}^0$ are linearly independent and any symmetric system is also
automatically hyperbolic.\newline
The previous technique was generalized in covariant relativistic formulation
by Ruggeri and Strumia \cite{Ruggeri-1981} that recognized the importance of
field that symmetrizes the original system and they proposed to call $%
\mathbf{u}^\prime$ \emph{main field}. Boillat called symmetric form (\ref%
{symm}) with relations (\ref{matrici}), \emph{Godunov systems}. This kind of systems are the typical ones of Rational  Extended Thermodynamics \cite{book}.

In the case of systems with parabolic structure (\emph{hyperbolic-parabolic
systems}), the following generalization of symmetric system (\ref{symm}) was
considered :
\begin{equation}
\mathbf{A}^0\,\frac{\partial \mathbf{u}^\prime}{\partial t}+\mathbf{A}%
^{j}\frac{\partial \mathbf{u}^\prime}{\partial x^{j}}
-\frac{\partial}{\partial
x^j} \left(\mathbf{B}^{jl}\frac{\partial \mathbf{u}^\prime}{\partial x^l }%
\right)=0,  \label{symmPara}
\end{equation}
where matrices\ $\mathbf{B}^{jl}= \left(\mathbf{B}^{jl}\right)^\star$\   are symmetric and $
\mathbf{B}_{(k)}=\mathbf{B}^{jl} k_j k_l $ \ are
non-negative definite.
\newline
The compatibility of system (\ref{symmPara}) with entropy principle and the
corresponding determination of main field was given by Ruggeri in \cite%
{Acta_Ruggeri} for Navier-Stokes-Fourier fluids and in general case by
Kawashima and Shizuta \cite{Kawa}. The same authors in \cite{Kawa2}
considered linearized version of system (\ref{symmPara}) proving that the
constant solutions are stable. For capillarity fluids, symmetric form (\ref%
{symmPara}) was studied in the simplest case by Gavrilyuk and Gouin \cite%
{Gavrilyuk2}.

Continuum models of capillarity can be interpreted by using
\emph{gradient theories} \cite{Germain1,Gouin3,Gavrilyuk}. The
models are useful to study interactions between fluids and solid
walls \cite{Gouin4,Gouin2} and they can be obtained thanks to
molecular methods \cite{Evans,Widom,Gouin1}. In fact, the
fourth-gradient model for fluids corresponds to   development  in
continuum mechanics when the principle of
  virtual powers  needs to obtain a separated form in
the sense of distributions' theory on the physical domain and its
boundaries, edges and end points where only vector forces are
applied at end points \cite{Schwartz,Gouin5}.

The study of models containing higher-order derivatives of the density has a clear interpretation in the framework of the mean-field molecular model.
In the mean-field theory of hard-sphere molecules, the van der Waals forces
exert stresses on fluid molecules producing surface tension effects \cite{Hamak,Lifshitz}. The second-gradient theory provides a construction of the energy density such that capillarity effects appear  as a consequence of the molecular model in domains where the change of mass density is important \cite{Widom2,rowlinson}.
\newline
The fourth-gradient model for fluids is the background of the paper:
the volume energy can be extended to obtain a fourth-gradient expansion of Cahn and Hilliard's equation \cite{Cahn} near the critical point \cite{Saccomandi}. The model is richer than the expansion of second order by van der Waals and others \cite{Waals}.
Such  extension obtained via the request of molecular range turns out to be effective in the construction of a new interpolating model compatible with fluctuations of density near the critical point; the effects are not negligible   and it is possible to deduce a Fisher-Kolmogorov equation \cite{Peletier} generating observable hydrodynamics fluctuations \cite{Swift}. The differences in pulse-wave oscillations between second- and fourth-gradient models allow to revisit papers introducing kinks versus pulses as in \cite{Truskinovsky}. We  believe that this result is remarkable and will hopefully stimulate further and deeper investigations on both theoretical and phenomenological nature. It is interesting to note -- and it
is not the case for the second-gradient model -- that the fourth gradient model is able to take the range of London intermolecular forces into account  \cite{Saccomandi}.

Using a statistical model in mean-field molecular theory, specific internal
energy $\varepsilon$ and volume free energy $F$ of the fourth-gradient fluid
are in the form,

\begin{equation}
\varepsilon = \alpha(\rho,s)- \frac{\lambda}{2}\, \Delta \rho - \frac{\gamma%
}{2}\, \Delta^2 \rho,  \label{graddeux}
\end{equation}
and
\begin{equation*}
F= f (\rho,T) - \frac{\lambda}{2}\,\rho\, \Delta\rho - \frac{\gamma}{2}\,
\rho\,\Delta^2 \rho,  \label{gradquatre}
\end{equation*}
with $\Delta\equiv {\rm div} \mathop{\rm grad}$ and $\Delta^2\equiv{\rm div}\big\{ \mathop{\rm grad}\, ({\rm div} \mathop{\rm grad})\big\} $ denote the harmonic and biharmonic operators,
where $\rm div$ and $\mathop{\rm grad} $ denote the divergence and gradient operators, respectively  ; $\rho$ is the fluid density, $s$ the specific entropy, $T$
the Kelvin temperature and $\lambda, \gamma$ are two scalar functions of $\rho$ and $s$ (or $\rho$ and $T$). Term $\alpha (\rho,s)$ is the specific internal
energy and $f (\rho,T)$ is the volume free energy of the homogeneous fluid
bulk of densities $\rho$ and $s$ at temperature $T$. In the mean-field
simplest  model, near the critical point of the fluid, $\lambda$ and $\gamma$
can be considered as constant, conditions  assumed along the paper.


\emph{\textbf{In case}} $\gamma = 0$,

We get the internal energy expression given in \cite{Cahn}. However, authors
used $\lambda/2\ (\mathop{\rm grad}\rho)^2$ in place of $- (\lambda / {%
2})\,\rho\, \Delta\rho$.\newline
But, $\rho\, \Delta \rho = \mathop{\rm div}(\rho\,\mathop{\rm grad}
\rho) - (\mathop{\rm grad} \rho)^2$; consequently, $\lambda / 2 \,
\mathop{\rm
div}(\rho\,\mathop{\rm grad} \rho)$ can be integrated on the fluid boundary
and is null when the fluid is homogeneous (as in the bulks).

\emph{\textbf{In case}} $\gamma \neq 0$,
\begin{equation*}
\rho\,\Delta^2 \rho = \mathop{\rm div}\big[\,\rho\,\mathop{\rm grad} (%
\mathop{\rm div}\mathop{\rm grad} \rho)- (\mathop{\rm div} \mathop{\rm grad}
\rho)\,\mathop{\rm grad} \rho\,\big]+[ \mathop{\rm div}\mathop{\rm grad}
\rho\,]^2.
\end{equation*}
Term $\mathop{\rm div}\big[\,\rho\,\mathop{\rm grad} (\mathop{\rm
div}\mathop{\rm grad} \rho)- (\mathop{\rm div} \mathop{\rm grad} \rho)\,%
\mathop{\rm grad} \rho\,\big]$ can be integrated on the boundary domain and
is null when the fluid is homogeneous (as in the bulks); then, $- ({\gamma}/{%
2})\,\rho\, \Delta^2 \rho$ can be replaced with $- \gamma / 2\,
(\Delta\rho\,)^2$.\newline
Consequently, for fourth-gradient fluids, the specific internal energy and
the free volume energy can be respectively replaced by:
\begin{equation}
\varepsilon = \alpha(\rho,s)+\frac{1}{\rho}\left( \frac{\lambda}{2}\, (%
\mathop{\rm grad} \rho)^2 - \frac{\gamma}{2}\, (\Delta \rho)^2\right),
\label{graddeux2}
\end{equation}
and
\begin{equation*}
F= f (\rho,T) +\frac{\lambda}{2}\, (\mathop{\rm grad} \rho)^2 - \frac{\gamma%
}{2}\, (\Delta \rho)^2,  \label{gradquatre2}
\end{equation*}

We note that the equation of motion is the same for the two energy
representations (\ref{graddeux}) and (\ref{graddeux2}) but the boundary
conditions, corresponding to the integrated terms, are different as it is
pointed out in \cite{Gouin1}.
  \newline

Here and later, for any vectors $\mathbf{a,b}$ we use the notation $%
\mathbf{a}^{\star }\mathbf{b}$ for the scalar product (the line is
multiplied by the column vector) and $\mathbf{a}^{\ }\mathbf{b}%
^{\star }$ for the tensor product (or $\mathbf{a}\otimes \mathbf{b}$
the column vector is multiplied by the line vector). Divergence of a linear
transformation $\mathbf{D}$ is the covector $\mathop{\rm div}(%
\mathbf{D})$ such that, for any constant vector $\mathbf{d}$, $%
\mathop{\rm div}(\mathbf{D})\, \mathbf{d}= \mathop{\rm div}(%
\mathbf{D}\,\mathbf{d})$. The identical transformation is denoted by
$\mathbf{I}$. \newline
The paper is organized as follows. In Section 2, thanks to the principle   of virtual powers, we obtain the equation of
conservative motions. In Section 3, we get the equation of energy and
extends the \emph{interstitial-working} notion obtained in second-gradient model \cite{Dunn}. In Section 4, we propose a system of
quasi-linear equations in divergence form. Using a convenient change of
variables associated with a Legendre transformation of the
total fluid energy,  near an equilibrium position we obtain an \emph{Hermitian
symmetric form} for the equations of perturbations. For the equations of \emph{fourth-gradient capillary fluids} that belong to the class of dispersive systems, we get an analog of symmetric form (\ref{symmPara}) with main field given by (\ref%
{mainfield})$^1$. The system is proved to be stable. A conclusion and two appendices end
the paper.

\section{Equation of conservative motions}

\subsection{The principle     of virtual
powers }

The principle of virtual powers is a convenient way to obtain the equation of
motions \cite{Lin,Seliger}. A particle is identified in Lagrange's
representation by a reference position $\mathbf{X}$ of coordinates $%
(X,Y,Z)$ belonging to reference configuration $\mathcal{D}_{0}$; its
position is given in physical space $\mathcal{D}$ by Euler's representation $%
\mathbf{x}$ of coordinates $(x,y,z)$. The variations of particle motions
are deduced from families of virtual motions of the fluid written as
\begin{equation*}
\mathbf{X}=\mathbf{\psi }(\mathbf{x},t;\beta ),
\end{equation*}%
where $\beta $ denotes a real parameter defined in the vicinity of $0$, and
the real motion corresponds to $\beta =0$. Virtual displacements in
reference configuration are associated with any variation of the real motion
written as in \cite{Gouin3},
\begin{equation*}
\tilde{\delta} \mathbf{X}=\left. \frac{\partial
\mathbf{\psi }}{\partial \beta }(\mathbf{x},t;\beta
)\right\vert _{\beta =0}.
\end{equation*}%
Variation $\tilde{\delta}$ is \textit{dual} and mathematically
equivalent to Serrin's variation denoted $\delta$ (\cite{Serrin},
p. 145).  It is important to note that - due to virtual
displacement $\tilde{\delta} \mathbf{X}$ - the variation
commutes with
the derivative with respect to physical-space variable $\mathbf{x}$ \ ($\tilde{\delta} \mathop{\rm grad}%
^{p}\rho =\mathop{\rm grad}^{p}\tilde{\delta} \rho ,\ p\in
\mathbb{N}$). Consequently, for complex fluids, $\tilde{\delta}$-variation is
straightforward and a lot simpler than ${\delta}$-variation
\cite{Gouin3,Gravgouin}.\newline Neglecting the body forces, the
Lagrangian of the fluid writes,
\begin{equation*}
L=\frac{1}{2}\,\rho \,{\mathbf{v}}^{\star }\mathbf{v}-\rho
\,\varepsilon ,
\end{equation*}%
where $\mathbf{v}$ denotes the particle velocity. Conservative
motions stationarize the Hamilton action
\begin{equation}
\mathcal{G}=\int_{\mathcal{D}}L~dx ,  \label{Hamilton}
\end{equation}
where   $dx $ denotes  the volume element in ${\mathcal{D}}$. The
density satisfies the mass conservation
\begin{equation}
\frac{\partial \rho }{\partial t}+\mathop{\rm div}(\rho \,\mathbf{v}%
)=0\qquad \ \ \ \Longleftrightarrow \qquad \rho \,\mathop{\rm det}%
\mathbf{F}=\rho _{_{0}}(\mathbf{X})  \label{density}
\end{equation}%
with $\mathbf{F}\equiv {\partial \mathbf{x}}/{\partial \mathbf{X}%
}$, where $\rho _{_{0}}$ is the reference density defined on $\mathcal{D}%
_{0} $. The specific entropy verifies
\begin{equation}
{\ \overset{{\ \centerdot }}{{\,s}}}=0\qquad \Longleftrightarrow \qquad
s=s_{_{0}}({\mathbf{X}})\,,  \label{entropy}
\end{equation}%
where $s_{_{0}}$ is defined on ${\mathcal{D}}_{0}$ and superposed dot
denotes the material derivative. \newline
Classical methods of variation calculus yield the variation of $\mathcal{G}$.
 Virtual
displacements can be assumed to be null in the vicinity of the boundary of $%
\mathcal{D}_{0}$ and consequently, variations of integrated terms are null
on the boundary $\mathcal{D}$. By using Stokes' formula, we can integrate by
parts the variations of integral (\ref{Hamilton}); from $\tilde{\delta} \mathcal{G}=%
\left(\partial\mathcal{G}(\beta )/\partial\beta\right)_{|{\beta =0}},$ we get (see Appendix A for
details)
\begin{equation*}
\tilde{\delta} \mathcal{G}=\int_{\mathcal{D}}\left\{\  \left[ \;\frac{1}{2}\,%
\mathbf{v}^{\star }\mathbf{v}-\rho \frac{\partial \alpha }{%
\partial \rho }-\alpha +\lambda \, {\Delta} \rho +\gamma \, {\Delta} ^{2}\rho \,
\right] \tilde{\delta} \rho -\rho \frac{\partial \alpha }{\partial
s}\,\tilde{\delta} s+\rho \,\mathbf{v}^{\star }\tilde{\delta}
\mathbf{v}\right\} ~dx.
\end{equation*}

\noindent Moreover:   \\

Equation (\ref{density}) implies
\begin{equation*}
\tilde{\delta} \rho =\rho \ {\mathop{\rm div}}_{0}\,\tilde{\delta} \mathbf{X}+\frac{1}{%
\det \mathbf{F}}\frac{\partial \rho _{_{0}}}{\partial \mathbf{X}}%
\,\tilde{\delta} \mathbf{X},\quad {\rm where} \quad
{\mathop{\rm div}}_{0}\,\tilde{\delta} \mathbf{X} =  {\rm
tr}\left(\frac{\partial\tilde{\delta}
\mathbf{X}}{\partial\mathbf{x}}
\frac{\partial\mathbf{x}}{\partial\mathbf{X}}\right)\equiv {\rm
tr}\left(\frac{\partial\tilde{\delta}
\mathbf{X}}{\partial\mathbf{X}}\right).
\end{equation*}%
Operator  $\ \mathop{\rm div}_{0}$\ denotes the divergence operator in $\mathcal{D}_{0}$.\\

The definition of the velocity implies
\begin{equation*}
\frac{\partial \mathbf{X}\,(\mathbf{x},t)}{\partial \mathbf{x}}\,%
\mathbf{v}+\frac{\partial \mathbf{X}\,(\mathbf{x},t)}{\partial t}%
=0,
\end{equation*}%
and consequently,
\begin{equation*}
\frac{\partial \tilde{\delta} \mathbf{X}}{\partial \mathbf{x}}\ \mathbf{v%
}+\frac{\partial \mathbf{X}}{\partial \mathbf{x}}\
\tilde{\delta} \mathbf{v}+\frac{\partial \tilde{\delta}
\mathbf{X}}{\partial t}=0\quad
\Longleftrightarrow \quad \tilde{\delta} \mathbf{v}=-F\overset{{\ \centerdot }}{%
\widehat{\tilde{\delta} \mathbf{X}}}.
\end{equation*}%
By denoting
\begin{equation*}
\quad H=\alpha +\frac{\mathcal{P}}{\rho }\ ,\quad K=H-\lambda \,
{\Delta}
\rho -\gamma \, {\Delta} ^{2}\rho \quad \mathrm{and}\quad m=\frac{1}{2}\,%
\mathbf{v}^{\star }\mathbf{v}-K\,,
\end{equation*}%
where $\mathcal{P}$ is the thermodynamical pressure, we obtain
\begin{equation*}
\tilde{\delta} \mathcal{G}=\int_{\mathcal{D}}\left[ m~\tilde{\delta} \rho -\rho \,{({%
\mathbf{v}^{\star }\mathbf{F}})}\,\overset{{\ \centerdot }}{\widehat{%
\tilde{\delta} \mathbf{X}}}-\rho \,T\left( {\mathop{\rm
grad}}_{_{0}}^{\star }\,s_{_{0}}\right) \tilde{\delta}
\mathbf{X}\right] ~dx \quad {\rm where} \quad {\mathop{\rm
grad}}_{0} s = \frac{\partial s_{_0}
({\mathbf{X}})}{\partial\mathbf{X}}
\end{equation*}%
and by integration by part on $\mathcal{D}_{0}$,
\begin{equation*}
\tilde{\delta} \mathcal{G}=\int_{{\mathcal{D}}_{0}}\rho _{0}\left[ \,\overset{%
\centerdot }{(\widehat{\mathbf{v}^{\star }\mathbf{F}})}-{\mathop{\rm
grad}}_{_{0}}^{\star }m-\,T\,{\mathop{\rm grad}}_{_{0}}^{\star }\,s_{_{0}}%
\right] \tilde{\delta} \mathbf{X}~dX.
\end{equation*}%
Terms  $\ \mathop{\rm grad}_{_{0}}$ and $dX$ denote the gradient
and the volume element in ${\mathcal{D}}_{0}$, respectively.

Due to the principle of virtual work : \newline

\centerline {\textit{For any displacement }$\tilde{\delta}
\mathbf{X}$\textit{\ null on the edge of }$\mathcal{D}_{0}$, $
\tilde{\delta} \mathcal{G}=0$,}

we get $\ \overset{\centerdot }{(\widehat{\mathbf{v}^{\star}\mathbf{F%
}})}= \mathop{\rm grad} _{_0}^{\star}\,m +T\  {\mathop{\rm grad}}
_{_0}^{\star}\, s_{_0}. $\newline
Noticing that  $\ (\mathbf{a}^{\star}+\displaystyle\mathbf{v}^{\star}%
\frac{\partial \mathbf{v}}{\partial \mathbf{x}})\mathbf{F}=%
\overset{\centerdot }{(\widehat{\mathbf{v}^{\star}\mathbf{F}})}$,
where $\mathbf{a}$ is the acceleration vector, we get
\begin{equation*}
\mathbf{a}+\mathop{\rm grad} K -T\, \mathop{\rm grad} s = 0.
\label{motion}
\end{equation*}%
But,
\begin{equation*}
dH = \frac{d\mathcal{P}}{\rho} + T\, ds
\end{equation*}
and consequently, the equation of motion writes
\begin{equation}
\rho\ \mathbf{a}+\mathop{\rm grad} \mathcal{P} -\lambda\,\rho\, %
\mathop{\rm grad} \Delta\rho - \gamma\,\rho\, \mathop{\rm grad}
\Delta^2\rho= 0 . \label{motion1}
\end{equation}

\subsection{Divergence form of the equation of motion}

On one hand, we note
\begin{equation*}
\mathbf{\sigma}_{_1} \equiv \lambda\left[\frac{1}{2} (\mathop{\rm grad}%
\rho)^2+\rho\,\Delta\rho\right] \mathbf{I} -\lambda\,(\mathop{\rm grad}%
\rho)\,{(\mathop{\rm grad}}^\star\rho)
\end{equation*}
Then,
\begin{equation*}
\mathop{\rm div}\mathbf{\sigma}_{_1}
=\lambda\left[{\mathop{\rm grad}}^\star\rho\,\frac{\partial
\mathop{\rm grad}
\rho}{\partial \mathbf{x}} +\Delta\rho\ {\mathop{\rm grad}}%
^\star\rho+\rho\  {\mathop{\rm grad}}^\star\Delta\rho- \Delta\rho\ {%
\mathop{\rm grad}}^\star\rho - {\mathop{\rm grad}}^\star\rho\,\frac{\partial %
\mathop{\rm grad} \rho}{\partial \mathbf{x}} \right],
\end{equation*}
and consequently,
\begin{equation*}
\mathop{\rm div}\mathbf{\sigma}_{_1} =\lambda\,\rho \, \mathop{\rm grad}%
\Delta\rho\,.
\end{equation*}
\qquad On the other hand,
\begin{equation}
\rho\, \mathop{\rm grad}\Delta^2\rho = \mathop{\rm
grad}(\rho\,\Delta^2\rho) - {\mathop{\rm
div}}^\star[\,(\mathop{\rm grad}\Delta\rho)\ {\mathop{\rm
grad}}^\star\rho\,] + \frac{\partial \mathop{\rm grad}
\rho}{\partial \mathbf{x}}\,\mathop{\rm grad} \Delta\rho
\label{Intermediate}
\end{equation}
and after some calculations (See Appendix B),
\begin{equation*}
 \mathop{\rm div}\mathbf{%
\sigma}_{_2}= \gamma\, \rho\, \mathop{\rm grad}\Delta^2\rho
\end{equation*}
with
\begin{equation*}
\mathbf{\sigma}_{_2}\equiv\gamma\left\{\ \left[\rho\,\Delta^2\rho-\frac{1}{2}%
\mathop{\rm tr}\left(\frac{\partial \mathop{\rm grad}
\rho}{\partial
\mathbf{x}}\right)^2\right] \mathbf{I} + \left(\frac{\partial %
\mathop{\rm grad} \rho}{\partial \mathbf{x}}\right)^2-
(\,\mathop{\rm grad}\Delta\rho)\,{\mathop{\rm
grad}}^\star\rho\right\}.
\end{equation*}
The equation of motion can be written in divergence form :
\begin{equation*}
\frac{\partial \rho\, \mathbf{v}^\star}{\partial t} + \mathop{\rm div}%
\left[\,\rho\, \mathbf{v} \mathbf{v}^\star+\mathcal{P}\,\mathbf{I%
}-\mathbf{\sigma}\right] = 0,
\end{equation*}
where
\begin{eqnarray*}
\mathbf{\sigma}&\equiv& \mathbf{\sigma}_{_1}+ \mathbf{\sigma}%
_{_2} \\
&=& \left[\frac{\lambda}{2}\, (\mathop{\rm grad}\rho)^2-\frac{\gamma}{2}%
\mathop{\rm tr}\,\left(\frac{\partial \mathop{\rm grad}
\rho}{\partial \mathbf{x}}\right)^2+\lambda\,\rho\,\Delta\rho+
\gamma\, \rho\,\Delta^2\rho \right] \mathbf{I}
 -\left(\lambda\,\mathop{\rm grad}\rho + \gamma \mathop{\rm grad}%
\Delta\rho\right)\, {\mathop{\rm grad}}^\star\rho + \gamma\,\left(\frac{%
\partial \mathop{\rm grad} \rho}{\partial \mathbf{x}}\right)^2
\end{eqnarray*}
In fact, $\,\mathbf{\sigma}$ has the physical dimension of a
stress tensor but is not a Cauchy stress tensor as we will notice
in  section 3.

\section{Equation of energy}

\noindent Multiplying Eq. (\ref{motion1}) by ${\mathbf{v}}$,
we get
\begin{equation*}
\rho\ \mathbf{a}^{\star}\,\mathbf{v}+(\mathop{\rm grad} \mathcal{P}%
)^{\star}\,\mathbf{v} -\lambda\,\rho\  (\mathop{\rm grad}
\Delta\rho) ^{\star}\,\mathbf{v} - \gamma\,\rho\, (\mathop{\rm
grad} \Delta^2\rho) ^{\star}\,\mathbf{v}= 0.
\label{energyscalaire}
\end{equation*}
Due to Gibbs' identity, the volume energy of the homogeneous fluid
yields
\begin{equation*}
   \rho\, T\, {\overset{ {\  \centerdot }}{s}} = \rho\, \frac{d\alpha}{dt } -\frac{\mathcal{P}}{\rho}\, {\overset{{\,\
\centerdot }}{\,\rho}}.
\end{equation*}
Taking eqs (\ref{density}) and (\ref{entropy}) into account, we
obtain
\begin{equation*}
\rho\,\frac{d}{dt}\left[\frac{1}{2}\,\mathbf{v}^2+\alpha-\lambda\,
\Delta\rho - \gamma\, \Delta^2\rho \right]+ \mathop{\rm div} \left(\mathcal{P%
}\mathbf{v}\right)+ \rho\,\Delta\left[\lambda\, \frac{\partial\rho}{%
\partial t}+\gamma\, \Delta\left(\frac{\partial\rho}{\partial t}\right)%
\right] = 0,
\end{equation*}
with $\mathbf{v}^2\equiv \mathbf{v}^\star \mathbf{v}\equiv |%
\mathbf{v}^2|$, and
\begin{eqnarray*}
&&\frac{\partial}{\partial t}\left[\rho\,\left( \frac{1}{2}\,\mathbf{v}%
^2+\alpha-\lambda\, \Delta\rho - \gamma\, \Delta^2\rho \right)\right]+ \\
&& \mathop{\rm div} \left[\rho\,\left( \frac{1}{2}\,\mathbf{v}%
^2+\alpha-\lambda\, \Delta\rho - \gamma\, \Delta^2\rho \right)\mathbf{v}+%
\mathcal{P}\mathbf{v}\right]+ \rho\,\Delta\left[\lambda \frac{%
\partial\rho}{\partial t}+\gamma \Delta\left(\frac{\partial\rho}{\partial t}%
\right)\right]=0.
\end{eqnarray*}
Taking account of relations
\begin{equation*}
- \lambda\,\Delta\rho\, \frac{\partial\rho}{\partial t}-\mathop{\rm div} %
\left[ \lambda\,\Delta\rho\ \rho\mathbf{v} \right] = \frac{\partial}{%
\partial t}\left[\frac{\lambda}{2} (\mathop{\rm grad}\rho)^2 \right]-%
\mathop{\rm div} \left(\lambda\,\Delta\rho\,\rho\mathbf{v} +
\lambda\, \frac{\partial\rho}{\partial t}\,\mathop{\rm
grad}\rho\right)
\end{equation*}
and
\begin{equation*}
 - \gamma\,\Delta^2\rho\, \frac{\partial\rho}{\partial t}-\mathop{\rm div} %
\left[ \gamma\,\Delta^2\rho\ \rho \mathbf{v}\right] =
 -\frac{\partial}{\partial t}\left(\frac{\gamma}{2} (\Delta \rho)^2 \right)+%
\mathop{\rm div} \left[\gamma\,\Delta\rho \,\mathop{\rm grad}\frac{%
\partial\rho}{\partial t} -\gamma\,\left(\frac{\partial\rho}{\partial t}%
\right)\,\mathop{\rm grad}\Delta \rho-\gamma\, \Delta^2 \rho\ \rho%
\mathbf{v}\right],
\end{equation*}
we obtain
\begin{eqnarray}
&&\frac{\partial}{\partial t}\left[\rho\,\left( \frac{1}{2}\,\mathbf{v}%
^2+\alpha\right)+\frac{\lambda}{2} (\mathop{\rm grad}\rho)^2- \frac{\gamma}{2%
} (\Delta \rho)^2 \right]+\mathop{\rm div} \left[\rho\,\left( \frac{1}{2}\,%
\mathbf{v}^2+\alpha-\lambda\, \Delta\rho - \gamma\, \Delta^2\rho \right)%
\mathbf{v}+\mathcal{P}\mathbf{v}\right]  \notag \\
&& - \mathop{\rm div} \left[\frac{\partial\rho}{\partial t}\left(\lambda\,%
\mathop{\rm grad}\rho+ \gamma\,\mathop{\rm grad}\Delta\rho\right)-\gamma\,%
\Delta\rho\,\mathop{\rm grad}\left(\frac{\partial\rho}{\partial t} \right) %
\right]=0.  \label{energy}
\end{eqnarray}
Equation (\ref{energy}) is the balance equation of energy of the
fourth-gradient fluid. Let us consider the specific energy in   form (\ref%
{graddeux2}), then the total volume energy of the fluid is,
\begin{equation}
e = \frac{1}{2}\,\rho\,\mathbf{v}^2+ \rho\,\alpha(\rho,s)+ \frac{\lambda%
}{2}\, (\mathop{\rm grad} \rho)^2 - \frac{\gamma}{2}\, (\Delta
\rho)^2 . \label{energy4}
\end{equation}
Term $\alpha + \mathcal{P}/\rho$ is the enthalpy of the
homogeneous bulk, and
\begin{equation*}
\mathcal{H} \equiv \rho\,\alpha + \frac{\mathcal{P}}{\rho} -
\lambda\, \Delta\rho - \gamma\, \Delta^2\rho
\end{equation*}
is the enthalpy of the fourth-gradient fluid. Let us note
\begin{equation*}
\Xi \equiv \frac{\partial\rho}{\partial t}\left(\lambda\,\mathop{\rm grad}%
\rho+ \gamma\,\mathop{\rm grad}\Delta\rho\right)-\gamma\,\Delta\rho\,%
\mathop{\rm grad}\left(\frac{\partial\rho}{\partial t} \right),
\end{equation*}
then, balance equation of energy (\ref{energy}) becomes
\begin{equation}
\frac{\partial e }{\partial t} + \mathop{\rm div}
\left[\left(\frac{1}{2}\, \mathbf{v}^2+
\mathcal{H}\right)\rho\,\mathbf{v} \right]-\mathop{\rm div}\Xi
= 0  .\label{energy2}
\end{equation}
In the special case of capillary fluids, Eq. (\ref{energy2})
reduces to
\begin{equation*}
\frac{\partial e_{_0} }{\partial t} + \mathop{\rm div} \left[ (e_{_0}-%
\mathbf{\sigma_{_1}})\mathbf{v} \right]-\mathop{\rm div}( \lambda\, {%
\overset{{\ \centerdot }}{\rho}}\, \mathop{\rm grad}\rho)= 0
\label{energy3}
\end{equation*}
where $\mathbf{\sigma}_{_1} = -p \,\mathbf{I} - \lambda\,
\mathop{\rm
grad}\rho\ \mathop{\rm grad}^\star\rho$ \   with\ $p = \mathcal{P}- \lambda\,(%
\mathop{\rm grad}\rho)^2/2 -\lambda\,\rho\,\Delta\rho$, \
corresponds to the
\emph{stress tensor}, $\lambda\, {\overset{{\ \centerdot }}{\rho}}\, %
\mathop{\rm grad}\rho$ is the \emph{interstitial working} vector and $%
\displaystyle e_{_0} = \frac{1}{2}\,\rho\,\mathbf{v}^2+\rho\,\alpha(%
\rho,s)+ \frac{\lambda}{2}\, (\mathop{\rm grad} \rho)^2$ is the
total volume energy of the capillary fluid, respectively. Or, with
\begin{equation*}
\quad \mathcal{H}_{_0} \equiv \rho\,\alpha +
\frac{\mathcal{P}}{\rho} -
\lambda\, \Delta\rho, \quad \mathrm{{and}\quad \Xi_{_0} \equiv \lambda\,%
\frac{\partial\rho}{\partial t} \,\mathop{\rm grad}\rho},
\end{equation*}
\begin{equation*}
\frac{\partial e_{_0} }{\partial t} + \mathop{\rm div} \left[\left(\frac{1}{2%
}\,\mathbf{v}^2+ \mathcal{H}_{_0}\right)\rho\,\mathbf{v} \right]-%
\mathop{\rm div} {\Xi}_{_0} = 0,
\end{equation*}
which is specific to gradient fluids because
$\mathbf{\sigma}_{_1}$ is not associated with a Cauchy stress
tensor of an elastic medium.

\section{Governing equations in symmetric form}

The internal energy per unit volume of the fourth-gradient fluid
is taken in the form
\begin{equation*}
\rho \,\varepsilon (\rho ,\eta ,\mathbf{w})=\epsilon (\rho ,\eta )+\frac{%
\lambda \left\vert \mathbf{w}\right\vert ^{2}}{2}-\frac{\gamma }{2}%
\,(\Delta \rho )^{2},
\end{equation*}%
where $\epsilon = \rho \,\alpha $,
$\mathbf{w}=\mathrm{{grad}\,\rho }\ $ and $\eta = \rho \,s$ is
the
entropy per unit volume. \emph{%
Homogeneous} internal energy per unit volume $\epsilon $\
satisfies the
Gibbs identity,%
\begin{equation*}
T\ d\eta =d\epsilon -\mu \ d\rho
\end{equation*}%
where $\mu =(\epsilon +\mathcal{P}-T\,\eta )/\rho \ $ is the
chemical potential of the fluid bulk. The governing equations of
the fourth-gradient
fluid write in the form%
\begin{equation}
\left\{
\begin{array}{l}
\quad \displaystyle\frac{\partial \rho }{\partial t}+\mathop{\rm div}%
\mathbf{j}=0 \\
\quad \displaystyle\frac{\partial \eta }{\partial t}+\mathop{\rm div}\left( %
\displaystyle\frac{\eta }{\rho }\ \mathbf{j}\right) =0\label%
{systemnhyperbolic1} \\
\quad \displaystyle\frac{\partial \mathbf{j}^{\star }}{\partial t}+%
\mathop{\rm div}\left( \displaystyle\frac{\mathbf{jj}^{\star }}{\rho }+%
\mathcal{P}\,\mathbf{I}\right) -\lambda \,\rho \ {\mathop{\rm grad}}%
^{\star }\left( \mathop{\rm div}\ \mathbf{w}\right) -\gamma \,\rho \,{%
\mathop{\rm grad}}^{\star }\Delta ^{2}\rho =\mathbf{0}^{\star }%
\end{array}%
\right.
\end{equation}%
where $\mathbf{j}\equiv \rho \,\mathbf{v}$. The gradient
of the mass
conservation law verifies another conservation law,%
\begin{equation}
\frac{\partial \mathbf{w}}{\partial t}+{\mathop{\rm
grad}}\mathop{\rm div}\mathbf{j}=0. \label{w}
\end{equation}%
Conversely, if we consider $\ \mathbf{w}\ $ as an independent
variable, and if we add the initial condition
\begin{equation*}
\mathbf{w}\,|_{\,t=0}={\mathop{\rm grad}}\rho \,|_{\,t=0}\,,
\end{equation*}%
$ \mathbf{w}=\mathrm{{grad}\,\rho }\, $ is a consequence of
the governing equations.
\newline
Similarly, we denote   $a=\Delta \rho\ $  the Laplace operator,
the mass conservation equation yields,
\begin{equation*}
\frac{\partial a}{\partial t}+\Delta (\,\mathop{\rm
div}\mathbf{j}\,)=0.
\end{equation*}%
Conversely, if we add the initial condition
\begin{equation*}
a\,|_{\,t=0}=\Delta \rho \,|_{\,t=0}\,,
\end{equation*}%
we can consider $\ a\ $ as an independent variable.\newline
Finally, we obtain the system of equations
(\ref{systemnhyperbolic1}) in the following equivalent
non-divergence form
\begin{equation}
\left\{
\begin{array}{l}
\quad \displaystyle\frac{\partial \rho }{\partial t}+\mathop{\rm div}%
\mathbf{j}=0 \\
\quad \displaystyle\frac{\partial \eta }{\partial t}+\mathop{\rm div}\left( %
\displaystyle\frac{\eta }{\rho }\ \mathbf{j}\right) =0\label%
{systemnhyperbolic2} \\
\quad \displaystyle\frac{\partial \mathbf{j}^{\star }}{\partial t}+%
\mathop{\rm div}\left( \displaystyle\frac{\mathbf{jj}^{\star }}{\rho }+%
\mathcal{P}\,\mathbf{I}\right) -\lambda \,\rho \ {\mathop{\rm grad}}%
^{\star }\left( \mathop{\rm div}\ \mathbf{w}\right) -\gamma \,\rho \,{%
\mathop{\rm grad}}^{\star }\Delta ^{2}\rho =\mathbf{0}^{\star } \\
\quad \displaystyle\frac{\partial \mathbf{w}}{\partial
t}+{\mathop{\rm
grad}}\mathop{\rm div}\mathbf{j}=0 \\
\quad \displaystyle\frac{\partial a}{\partial t}+\Delta (\,\mathop{\rm div}%
\mathbf{j}\,)=0%
\end{array}%
\right.
\end{equation}%
{\textbf{Remark}:} We choose energy equation (\ref{energy}) as
supplementary equation. In usual thermodynamical theories the
energy equation is a part of the system and the entropy balance
equation is taken as a supplementary equation (entropy principle).
In the case of weak solutions, the fact is very important; in
particular, for shock waves, the  entropy is growing across the
shock. But when we consider classical solutions, we can, without
losing generality, switch   roles of entropy and energy.

The theory of capillary usually applied for van der Waals-like
fluids can be extended to fourth-gradient fluids. For such fluids
the energy $\epsilon
\left( \rho ,\eta \right) $ is not convex for all values of $\rho $ and $%
\eta $. We assume that we are in the vicinity of an equilibrium state $%
\left( \rho _{e},\eta _{e}\right) $ where the energy function is
locally convex. \newline
With $\mathbf{u}\equiv (\rho ,\eta ,\mathbf{j}%
^{\star },\mathbf{w}^{\star },a)^{\star }$ and \ $h^{0}\equiv
e$\ \ (given by Eq. (\ref{energy4})),  from Eq.
(\ref{mainfield})$^{1}$ we deduce the main field
\begin{equation*}
\mathbf{u}^{\prime }\equiv (q,\ \theta ,\
\mathbf{v}^{\prime ^{\star }},\ \mathbf{r}^{\star },\ b\
)^{\star }
\end{equation*}%
coming from
\begin{equation*}
\begin{array}{l}
de = \displaystyle\left( \mu -\frac{\left\vert
\mathbf{v}\right\vert
^{2}}{2}\right) d\rho +T\,d\eta +\mathbf{v}^{\star }d\mathbf{j}%
+\lambda \,\mathbf{w}^{\star }d\mathbf{w}-\gamma \,a\,da \\
\ \ \quad =\displaystyle q\,d\rho +\theta \,d\eta
++\mathbf{v}^{\prime
\star }d\mathbf{j}+\mathbf{r}^{\star }d\mathbf{w}+b\,da%
\end{array}%
\end{equation*}%
and therefore
\begin{equation*}
q=\mu -\frac{\left\vert \mathbf{v}\right\vert ^{2}}{2},\quad
\theta =T,\quad \mathbf{v}^{\prime }=\mathbf{v},\quad
\mathbf{r}^{\star }=\lambda \,{\mathop{\rm grad}}^{\star }\rho
\quad \mathrm{and}\quad b=-\gamma \,\Delta \rho \ .
\end{equation*}%
 Legendre transformation $h^{\prime 0}=\Pi $ of   total energy $h^{0}=e$
given by Eq. (\ref{mainfield})$^{2}$ is
\begin{equation*}
{\Pi }=\rho \,q+\eta \,T+\mathbf{j}^{\star }\mathbf{v}+\mathbf{w}%
^{\star }\mathbf{r}+a\,b-E={\mathcal{P}}+\frac{\left\vert \mathbf{r}%
\right\vert ^{2}}{2\,\lambda }-\frac{b^{2}}{2\,\gamma },
\end{equation*}%
where   thermodynamic pressure $\mathcal{P}$ is considered as a
function
of $\ q,\ \theta $ and $\mathbf{v}$. Therefore, from Eq. (\ref{change})$%
^{1}$ we get
\begin{equation*}
\frac{\partial \Pi }{\partial q}=\rho ,\quad \frac{\partial \Pi }{\partial T}%
=\eta ,\quad \frac{\partial \Pi }{\partial \mathbf{u}}=\mathbf{j}%
^{\star },\quad \frac{\partial \Pi }{\partial \mathbf{r}}=\mathbf{w}%
^{\star },\quad \frac{\partial \Pi }{\partial b}=a,
\end{equation*}%
If we introduce matrix $\displaystyle\mathbf{B}\equiv -\gamma \,\frac{%
\partial \mathop{\rm grad}\rho }{\partial \mathbf{x}},\ $ System (\ref%
{systemnhyperbolic2}) can be rewritten as a symmetric form (\ref{symmPara}%
) in which the hyperbolic part is in the form (\ref{symform}) ,
(\ref{symm}) :

\begin{equation}
\left\{
\begin{array}{l}
\quad \displaystyle\frac{\partial }{\partial t}\left( \frac{\partial \Pi }{%
\partial q}\right) +\mathop{\rm div}\left[ \frac{\partial (\Pi \mathbf{v)%
}}{\partial q}\right] =0 \\
\quad \displaystyle\frac{\partial }{\partial t}\left( \frac{\partial \Pi }{%
\partial T}\right) +\mathop{\rm div}\left[ \frac{\partial (\Pi \mathbf{v)%
}}{\partial T}\right] =0\label{system4} \\
\displaystyle\quad \frac{\partial }{\partial t}\left( \frac{\partial \Pi }{%
\partial \mathbf{v}}\right) +\mathop{\rm div}\left[ \frac{\partial (\Pi
\mathbf{v)}}{\partial \mathbf{v}}-\frac{\partial \Pi }{\partial q}%
\dfrac{\partial \mathbf{r}}{\partial \mathbf{x}}+\left\{ \frac{1}{2}%
\frac{\partial \Pi }{\partial b}\,b-\frac{1}{2}\mathop{\rm
tr}\left( \mathbf{B}\frac{\partial }{\partial
\mathbf{x}}\left( \frac{\partial }{\partial
\mathbf{x}}\frac{\partial \Pi }{\partial q}\right) ^{\star
}\right) \right. \right. \\
\displaystyle\left. \left. \quad \displaystyle+\frac{\partial \Pi
}{\partial q}\mathop{\rm tr}\left( \frac{\partial }{\partial
\mathbf{x}}\left(
\frac{\partial }{\partial \mathbf{x}}\frac{\partial \Pi }{\partial b}%
\right) ^{\star }\right) \right\} \mathbf{I}-(\mathop{\rm grad}b)\ {%
\mathop{\rm grad}}^{\star }\left( \frac{\partial \Pi }{\partial q}\right) +%
\mathbf{B}\,\frac{\partial }{\partial \mathbf{x}}\left( \frac{%
\partial (\frac{\partial \Pi }{\partial q})}{\partial \mathbf{x}}\right)
^{\star }\right] =0 \\
\displaystyle\quad \frac{\partial }{\partial t}\left( \frac{\partial \Pi }{%
\partial \mathbf{r}}\right) +\mathop{\rm div}\left[ \frac{\partial (\Pi
\mathbf{v)}}{\partial \mathbf{r}}+\frac{\partial \Pi }{\partial q}%
\dfrac{\partial \mathbf{v}}{\partial \mathbf{x}}\right] =0 \\
\displaystyle\quad \frac{\partial }{\partial t}\left( \frac{\partial \Pi }{%
\partial b}\right) +\mathop{\rm div}\left[ \,\frac{\partial }{\partial
\mathbf{x}}\left( \mathop{\rm tr}\left\{ \frac{\partial
}{\partial \mathbf{x}}\left( \frac{\partial (\Pi }{\partial
\mathbf{v}}\right)
\right\}\, \right) ^{\star }\right] ^{\star }=0,%
\end{array}%
\right.
\end{equation}

Therefore, the system has a Cauchy problem well posed according with the general results proved in \cite{Kawa,Kawa2} for hyperbolic-parabolic systems in  form (\ref{symmPara}). If the capillary coefficients $\lambda$ and $\gamma$ are zero, $%
\Pi =\mathcal{P}$ and we get   gas-dynamics' equation and the
symmetric hyperbolic form of Godunov \cite{Godunov}.

\section{Stability of constant states}

System (\ref{system4}) admits constant solutions $(\rho _{e},\eta _{e},%
\mathbf{v}_{e},\mathbf{w}_{e}=\mathbf{0}, a_e = 0)$.\
Since the governing equations are invariant under Galilean
transformation, we can assume that
$\mathbf{v}_{e}=\mathbf{0}$.

Near equilibrium, we look for the solutions of the linearized
system
proportional to $\displaystyle e^{i\left( \mathbf{k}^{\star }\mathbf{%
x}-\omega t\right) },\ (i^2 =-1 , \ \mathbf{k}^{\star }
\mathbf{k} =1) :$
\begin{equation*}
\mathbf{u} = \mathbf{u}_{_0}e^{i\left( \mathbf{k}^{\star }%
\mathbf{x}-\omega t\right) } \quad \mathrm{with}\quad\mathbf{u}^\star = %
\left[\,q, T, \mathbf{v}, \mathbf{r}, b\,\right]\quad
\mathrm{and} \quad \mathbf{u}_{_0}^\star = \left[\,q_{_0},
T_{_0}, \mathbf{u}_{_0}, \mathbf{r}_{_0}, b_{_0}\,\right].
\label{perturbation}
\end{equation*}
We obtain
\begin{equation*}
\frac{\partial}{\partial t}\left( \frac{\partial \Pi }{\partial \mathbf{u%
}}\right)_{e} = \frac{\partial}{\partial \mathbf{u}}\left( \frac{%
\partial \Pi }{\partial \mathbf{u}}\right)_{e}\frac{\partial \mathbf{%
u}}{\partial t} = -i\,\omega \,\frac{\partial}{\partial \mathbf{u}}%
\left( \frac{\partial \Pi }{\partial \mathbf{u}}\right)_{e} \mathbf{u%
}_{_0}\, e^{i\left( \mathbf{k}^{\star }\mathbf{x}-\omega
t\right) }
\end{equation*}
where subscript $e$ means at equilibrium and we note $\mathbf{m}^\star = %
\left[\,q, T, \mathbf{v}, \mathbf{r}\,\right]$ and $\mathbf{m}%
_{_0}^\star = \left[\,q_{_0}, T_{_0}, \mathbf{v}_{_0}, \mathbf{r}%
_{_0}\,\right]$ such that $\mathbf{u}^\star =\left[\,\mathbf{m}%
^\star, b\,\right ]$ and $\mathbf{u}_{_0}^\star =\left[\,\mathbf{m}%
_{_0}^\star, b_{_0}\,\right ]$ .
\begin{equation*}
\text{div}\left(\frac{\partial\, \Pi \mathbf{v} }{\partial \mathbf{m}%
}\right)= \sum_{j=1}^3 \left[\left(\frac{\partial\, \Pi v^j }{\partial \mathbf{%
m}}\right),_{{x^j}}\right]^\star= \sum_{j=1}^3
\frac{\partial}{\partial
\mathbf{m}}\left(\frac{\partial\, \Pi v^j }{\partial \mathbf{m}}%
\right)^\star \frac{\partial\mathbf{m}}{\partial{x^j}},
\end{equation*}
and at equilibrium,
\begin{equation*}
\text{div}\left(\frac{\partial\, \Pi \mathbf{v} }{\partial \mathbf{m}%
}\right)_e = \sum_{j=1}^3
i\,F^j\,k_j\,\mathbf{m}_{_0}\,e^{i\left( \mathbf{k}^{\star
}\mathbf{x}-\omega t\right) },
\end{equation*}
where
\begin{equation*}
F^j \equiv \frac{\partial}{\partial
\mathbf{m}}\left(\frac{\partial\, \Pi v^j }{\partial
\mathbf{m}}\right)^\star_{e} \quad \mathrm{and\ we \
note}\quad {F} \equiv \sum_{j=1}^3 F^j\, k_j .
\end{equation*}

\quad $\bullet$ To Eq. (\ref{systemnhyperbolic2})$^3$ (or equivalently Eq. (%
\ref{system4})$^3$), we must add two   terms with respect to
classical fluids' equations :
\\
\emph{\ First term,}
\begin{equation*}
-\lambda \,\rho \ {\mathop{\rm grad}}^{\star }(\text{div}\,\mathbf{w}%
) = - \rho\, \text{div}
\left(\frac{\partial\mathbf{r}}{\partial
\mathbf{x}}\right) \quad \mathrm{with}\quad \frac{\partial\mathbf{r}%
}{\partial \mathbf{x}} = i \,
\mathbf{r}_{_0}\mathbf{k}^\star e^{i\left(
\mathbf{k}^{\star } \mathbf{x}-\omega t\right)},
\end{equation*}
where $\mathbf{r}=\mathbf{r}_{_0} e^{i\left(
\mathbf{k}^{\star } \mathbf{x}-\omega t\right)}$;
consequently,
\begin{equation*}
[\,-\lambda \,\rho \ {\mathop{\rm grad}}^{\star }(\text{div}\,\mathbf{w}%
)\,]_e = \rho_e \,\mathbf{r}_{_0}^{\star }\, \mathbf{k} \,
\mathbf{k}^{\star }\, e^{i\left( \mathbf{k}^{\star } \mathbf{x}%
-\omega t\right)}.
\end{equation*}
\emph{Second term,}
\begin{equation*}
-\gamma \,\rho \,{\mathop{\rm grad}}^{\star }\Delta ^{2}\rho = \rho \ {%
\mathop{\rm grad}}^{\star }(\text{div}\,\text{grad}\, b) \quad \mathrm{with}%
\quad b = b_{_0}\, e^{i\left( \mathbf{k}^{\star }
\mathbf{x}-\omega t\right)} .
\end{equation*}
But
\begin{equation*}
-\gamma\,\Delta ^{2}\rho = \text{div}\,\text{grad}\, b =
b_{_0}\,i^2\, \mathbf{k}^{\star }   \mathbf{k}\ e^{i\left(
\mathbf{k}^{\star } \mathbf{x}-\omega t\right)} .
\end{equation*}
Then, at equilibrium,
\begin{equation*}
-\gamma \,\rho \,{\mathop{\rm grad}}^{\star }\Delta ^{2}\rho =
-\,i\,\rho_e\, b_{_0} \, e^{i\left( \mathbf{k}^{\star }
\mathbf{x}-\omega t\right)}
  \mathbf{k}^\star.
\end{equation*}

\quad $\bullet$ To Eq. (\ref{system4})$^4$  at equilibrium, we
must add  the term,
\begin{equation*}
\mathop{\rm div}\left[\frac{\partial \Pi }{\partial
q}\dfrac{\partial
\mathbf{v}}{\partial \mathbf{x}}\right]_e = -\rho_e\, \mathbf{v}%
_{_0}^{\star } \mathbf{k}\,\mathbf{k}^{\star }e^{i\left( \mathbf{%
k}^{\star } \mathbf{x}-\omega t\right)} \quad
\mathrm{with}\quad \mathbf{v} = \mathbf{v}_{_0}\,
e^{i\left( \mathbf{k}^{\star } \mathbf{x}-\omega t\right)}
.
\end{equation*}

\quad $\bullet$ To Eq. (\ref{systemnhyperbolic2})$^5$ (or equivalently Eq. (%
\ref{system4})$^5$), we must add the term,
\begin{equation*}
\begin{array}{l}
\Delta (\,\mathop{\text{div}}\mathbf{j}\,) = \text{div}\text{grad}%
\left[({\text{grad}}\, \rho)^\star\ \mathbf{v}+ \rho\,
\text{div}
\mathbf{v}\right ]= \\
\text{div}\displaystyle\left[\left(\frac{\partial
\text{grad}\rho}{\partial
\mathbf{x}}\right)^\star\right]\ \mathbf{v}+ \text{tr}\left[\left(%
\frac{\partial \text{grad}\rho}{\partial \mathbf{x}}\right)^\star\frac{%
\partial \mathbf{v}}{\partial \mathbf{x}}+\left(\frac{\partial
\mathbf{v}}{\partial \mathbf{x}}\right)^\star\frac{\partial \text{%
grad}\rho}{\partial \mathbf{x}}\right] + \text %
{div}\left[\left(\frac{\partial
\mathbf{v}}{\partial \mathbf{x}}\right)^\star\right]\ \text{grad}\rho \\
+\displaystyle(\text{div} \mathbf{v})\, ( \text{div}\text{grad}\rho )+%
\frac{\partial\text{div}\mathbf{v}}{\partial\mathbf{x}}\,\text{grad}
\rho+ \rho\,\Delta (\text{div}\mathbf{v})+
{\text{grad}}^\star\rho\,
\text{grad}(\text{div}\mathbf{v}) .%
\end{array}%
\end{equation*}
At equilibrium, near $\rho = \rho_e$, the only remaining term is\ $%
\rho_e\,\Delta (\text{div}\mathbf{v})$, and taking
$\mathbf{v} =
\mathbf{v}_{_0}\, e^{i\left( \mathbf{k}^{\star } \mathbf{x}%
-\omega t\right)}$ into account, we obtain
\begin{equation*}
\rho_e\,\Delta (\text{div}\mathbf{v}) = -\, i \, \rho_e\, \mathbf{k}%
^\star  \mathbf{v}_{_0}\, e^{i\left( \mathbf{k}^{\star }
\mathbf{x}-\omega t\right)}.
\end{equation*}
Let us denote
\begin{equation*}
\begin{array}{c}
\displaystyle \mathbf{A}=\frac{\partial }{\partial
\mathbf{u}}\left[
\left( \frac{\partial \Pi }{\partial \mathbf{u}}\right) ^{\star }\right]%
_e ,\quad \mathbf{G}= \left[
\begin{array}{cc}
F & \underline{\mathbf{0}} \\
\underline{\mathbf{0}}^{\star} & 0%
\end{array}%
\right],%
\end{array}%
\end{equation*}
where $\underline{\mathbf{0}}$ and
$\underline{\mathbf{0}}^{\star}$
are column and raw matrices with nine zeros: $\underline{\mathbf{0}}%
^{\star} =[\,0\ 0\ 0\ 0\ 0\ 0\ 0\ 0\ 0\,]$,
\begin{equation*}
\mathbf{H}= \,\rho_e\, \left[
\begin{array}{ccccc}
0 & 0 & \mathbf{0}^{\star } & \mathbf{0}^{\star } & 0 \\
0 & 0 & \mathbf{0}^{\star } & \mathbf{0}^{\star } & 0 \\
\mathbf{0} & \mathbf{0} & \mathbf{0}_3 & - i\, \mathbf{kk}%
^{\star } & -\mathbf{k} \\
\mathbf{0} & \mathbf{0} & i\, \mathbf{kk}^{\star } &
\mathbf{0}_3  &
\mathbf{0} \\
0 & 0 & -\mathbf{k}^{\star } & \mathbf{0}^{\star } & 0%
\end{array}%
\right] \quad \text{with}\quad \mathbf{0}_3=\left[
\begin{array}{ccc}
0 & 0 & 0 \\
0 & 0 & 0 \\
0 & 0 & 0%
\end{array}%
\right] \quad \text{and}\quad \mathbf{0}^{\star }=[\,0\ 0\
0\,] .
\end{equation*}
Due to $\overline{\mathbf{H}}^{\;\star} = \mathbf{H}$,
matrices $\mathbf{G}$ and $\mathbf{H}$ are Hermitian and
the perturbations of system (\ref{system4}) verify
\begin{equation*}
i\left[\,\mathbf{C} - \omega\mathbf{A}\,\right]\, \mathbf{u}%
_{_0}\, e^{i\left( \mathbf{k}^{\star }\mathbf{x}-\omega
t\right) } = \mathbf{0},
\end{equation*}
where $\mathbf{C} \equiv \mathbf{G}+\mathbf{H}$\ and $%
\mathbf{A}$ are Hermitian and symmetric matrices,
respectively. Consequently, $\omega$-values are the roots of the
characteristic equation
\begin{equation*}
\det\left[\,\mathbf{C} -\omega\,\mathbf{A}\,\right] = {0},
\end{equation*}
where $\omega$ are the eigenvalues of $\mathbf{C}$ with respect to $%
\mathbf{A}$ and $\mathbf{u}_{_0}$ are the corresponding
eigenvectors. Hence, $\omega$ is real if $\mathbf{A}$ is
positive definite.

\section{Conclusion}

The fourth-gradient model of capillarity yields a conservation
energy equation. By a Legendre transformation of energy variables,
its quasi-linear system of conservation laws can be symmetrized in
the sense of Hermitian matrices.

This result extends the simplest case of capillarity with
second-gradient model \cite{Gavrilyuk2} and the problem of
stability of fluids in gradient theories for mass density.

\appendix

\bigskip

\section{Useful formulae}

\begin{equation*}
\rho\ \mathrm{{div} (\mathop{\rm grad} \rho)= {div}
(\rho\mathop{\rm grad} \rho) -(\mathop{\rm grad} \rho)^2.}
\end{equation*}
Term $\mathrm{{div} (\rho\mathop{\rm grad} \rho)}$ can be
integrated on the
boundary of $\mathcal{D}$ and consequently $\displaystyle -\,\tilde{\delta}\left(%
\frac{\lambda}{2}\, (\mathop{\rm grad} \rho)^2\right)$ corresponds in $%
\mathcal{D}$ to
\begin{equation*}
-\lambda \,{\mathop{\rm grad}}^\star \rho\ \mathop{\rm grad}
\tilde{\delta}\rho = -\lambda\, \mathrm{{div} (\tilde{\delta}\rho
\mathop{\rm grad} \rho) + \lambda\, {div} (\mathop{\rm grad}
\rho)\,\tilde{\delta}\rho }.
\end{equation*}
Term $\mathrm{div} (\tilde{\delta}\rho \mathop{\rm grad} \rho)$
can be integrated on
the boundary of $\mathcal{D}$ and the variation of $\displaystyle \frac{%
\lambda}{2}\,\rho\,\Delta\rho$ is $\lambda\;\Delta\rho\
\tilde{\delta}\rho$.\newline

In a similar way,
\begin{equation*}
\rho \ \mathrm{{div}\big[\mathop{\rm grad}({div}\mathop{\rm grad}\rho )\big]%
= {div}\big[\rho \mathop{\rm grad}({div}\mathop{\rm grad}\rho )\big]-{%
\mathop{\rm grad}}^{\star }\rho \ \mathop{\rm grad}({div}\mathop{\rm grad}%
\rho )}.
\end{equation*}%
but, $\mathrm{{div}\big[\rho \mathop{\rm grad}({div}\mathop{\rm
grad}\rho )\big]}$ can be integrated on the boundary of
$\mathcal{D}$ and
\begin{equation*}
-{\mathop{\rm grad}}^{\star }\rho \ \mathop{\rm grad}(\mathrm{{div}%
\mathop{\rm grad}\rho ) = -{div}\big[({div} \mathop{\rm grad}\rho )%
\mathop{\rm grad}\rho \big]+\big({div} \mathop{\rm grad}\rho
\big)^{2}.}
\end{equation*}%
Integrating on the boundary of $\mathcal{D}$\, term\, $-\mathrm{{div}\big[({div}%
\mathop{\rm grad}\rho )\mathop{\rm grad}\rho \big]}$, and
considering that
variation of $\rho \ \mathrm{{div}\big[\mathop{\rm grad}({div}%
\mathop{\rm
grad}\rho )\big]}$ is the same as variation of $\big(\mathrm{{div} %
\mathop{\rm grad}\rho \big)^{2}}$, we obtain
\begin{equation*}
2\ (\mathrm{div}\, \mathrm{grad}\,\,\rho )\ (\mathrm{div}\,
\mathrm{grad}\,\tilde{\delta} \rho ) = 2\;\mathrm{div}\big[
(\mathrm{{div}\,grad\,\rho)\,grad\,\tilde{\delta} \rho \big]-2\
{grad}^{\star }(div\,grad\,\rho )\ grad\,\tilde{\delta} \rho .}
\end{equation*}%
Term $2\;\mathrm{{div}\big[({div}\mathop{\rm grad}\rho
)\,\mathop{\rm grad}\tilde{\delta}\rho \big]}$ can be integrated
on the boundary of $\mathcal{D}$ and
\begin{equation*}
-2\,\mathrm{grad}^{\star }({\mathrm{d}iv}\,\mathrm{grad}\rho )\,\mathrm{grad}%
\,\tilde{\delta}\rho = -2\,\mathrm{div}\big[\tilde{\delta} \rho \ \mathrm{grad}(%
\mathrm{div}\,\mathrm{grad}\rho )\big]+2\,\left[\mathrm{div}\,\mathrm{grad}(%
\mathrm{div}\,\mathrm{grad}\rho )\right]\,\tilde{\delta} \rho .
\end{equation*}%
Term $-2\,\mathrm{{div}\big[ \tilde{\delta} \rho \,\mathop{\rm
grad}({div}\mathop{\rm grad}\rho )\big]}$ can be integrated on the
boundary of $\mathcal{D}$ and variation of
$\displaystyle\frac{\gamma }{2}\,\rho \,\Delta ^{2}\rho $ is
\begin{equation*}
\gamma \,[\,\mathrm{{div}\mathop{\rm grad}({div}\mathop{\rm
grad}\rho )\,]\,\tilde{\delta} \rho = \gamma \,(\Delta ^{2}\rho
)\,\tilde{\delta} \rho \,.}
\end{equation*}

\section{Additive calculations  to Subsection 2.2}

 In Rel. (\ref{Intermediate})   we have to study term  $\displaystyle
\frac{\partial \mathop{\rm grad} \rho}{\partial \mathbf{x}}\,%
\mathop{\rm grad}( \mathop{\rm div}\mathop{\rm grad}\rho)$. Due to
\begin{equation*}
\mathop{\rm grad}( \mathop{\rm div}\mathop{\rm grad}\rho) = {\mathop{\rm div}%
}^\star\left( \frac{\partial \mathop{\rm grad} \rho}{\partial \mathbf{x}}%
\right)\quad \mathrm{and } \quad \frac{\partial \mathop{\rm grad} \rho}{%
\partial \mathbf{x}} = \left( \frac{\partial \mathop{\rm grad} \rho}{%
\partial \mathbf{x}}\right)^\star,
\end{equation*}
\begin{equation*}
\frac{\partial \mathop{\rm grad} \rho}{\partial
\mathbf{x}}\,\mathop{\rm
grad}( \mathop{\rm div}\mathop{\rm grad}\rho) =\left[ \mathop{\rm div}\left(%
\frac{\partial \mathop{\rm grad} \rho}{\partial \mathbf{x}}\right) \frac{%
\partial \mathop{\rm grad} \rho}{\partial \mathbf{x}}\right]^\star
\end{equation*}
Each term of   covector \quad $\displaystyle \mathop{\rm div}\left(\frac{%
\partial \mathop{\rm grad} \rho}{\partial \mathbf{x}}\right)\ \frac{%
\partial \mathop{\rm grad} \rho}{\partial \mathbf{x}}$\quad is in the
form $\left\{\rho,_{kkj} \rho,_{jl}\right\}$.\newline From
\begin{equation*}
\rho,_{kkj}\rho,_{jl} = \left(\rho,_{lj}\rho,_{jk}\right),_k -
\rho,_{ljk}\rho,_{jk}
\end{equation*}
and
\begin{equation*}
\left(\rho,_{jk}\rho,_{jk}\right),_l= \rho,_{jkl}\rho,_{jk}+
\rho,_{jk}\rho,_{jkl},
\end{equation*}
together with the Schwarz theorem we get
\begin{equation*}
\rho,_{ljk}\rho,_{jk} =
\frac{1}{2}\left(\rho,_{jk}\rho,_{jk}\right),_l
\end{equation*}
which are the elements of $\displaystyle\frac{1}{2}\,\frac{\partial}{%
\partial \mathbf{x}}\left[\mathop{\rm tr}\left(\frac{\partial %
\mathop{\rm grad} \rho}{\partial \mathbf{x}}\right)^2\right]$. But $%
\left(\rho,_{lj}\rho,_{jk}\right),_k $ are the elements of $\displaystyle%
\mathop{\rm div}\left(\frac{\partial \mathop{\rm grad}
\rho}{\partial \mathbf{x}}\right)^2$. Consequently,
\begin{equation*}
\frac{\partial \mathop{\rm grad} \rho}{\partial
\mathbf{x}}\,\mathop{\rm
grad}( \mathop{\rm div}\mathop{\rm grad}\rho) = {\mathop{\rm div}}%
^\star\left(\frac{\partial \mathop{\rm grad} \rho}{\partial \mathbf{x}}%
\right)^2 - \frac{1}{2}\,\mathop{\rm grad}\left[\mathop{\rm tr}\left(\frac{%
\partial \mathop{\rm grad} \rho}{\partial \mathbf{x}}\right)^2\right].
\end{equation*}

\bigskip

\parindent 0pt {\small {\textbf{Acknowledgments}: This work was supported in
part (H.G.) by Institut Carnot and in part (T.R.) by National
Group of Mathematical Physics GNFM-INdAM.} }

\end{document}